\documentclass{article}

\usepackage{authblk}
\usepackage{amsmath}
\usepackage{amsfonts}
\usepackage{graphicx,subcaption}
\usepackage{url}
\title{Labyrinth walks: An elegant chaotic conservative non-Hamiltonian system}

\author[1]{Anouchah Latifi}
\author[2,3]{Vasileios Basios}
\author[4]{Chris G. Antonopoulos}

\affil[1]{Department of Physics, Faculty of Sciences, Qom University of Technology, Iran\thanks{E-mail: latifi@qut.ac.ir}}

\affil[2]{Service de Physique des Syst\`emes Complexes et M\'ecanique Statistique, Universit\'e Libre de Bruxelles, Belgium}

\affil[3]{Interdisciplinary Center for Nonlinear Phenomena and Complex Systems (CeNoLi), Belgium\thanks{E-mail: vbasios@ulb.ac.be}}

\affil[4]{Department of Mathematical Sciences, University of Essex, UK\thanks{E-mail: canton@essex.ac.uk (corresponding author)}}

\date{\today}

\begin{document}

\maketitle
\begin{abstract}
In this paper, we show that ``Labyrinth walks'', the conservative version of ``Labyrinth chaos'' and member of the Thomas-R\"ossler class of systems, does not admit an autonomous Hamiltonian as a constant function in time, and as a consequence, does not admit a symplectic structure. However, it is conservative, and thus admits a vector potential, being at the same time chaotic. This exceptional set of properties makes ``Labyrinth walks'' an elegant example of a chaotic, conservative, non-Hamiltonian system, with only unstable stationary points in its phase space, arranged in a 3-dimensional grid. As a consequence, ``Labyrinth walks'', even though is a deterministic system, it exhibits motion reminiscent of fractional Brownian motion in stochastic systems!
\end{abstract}

\vskip 1cm

\vskip 1cm

\section{Introduction}

The investigations on the Thomas-R\"ossler (TR) class of systems were initiated by R. Thomas and O. R\"{o}ssler's pioneering work \cite{Thomas1999, ThomRoss2004, ThomRoss2006}, while they were examining the role of feedback circuits and their related logical (Boolean) structure on the necessary conditions for the appearance of chaos. The biological relevance of this approach has been a very active and fruitful area of research \cite{Kaufman2013, Kaufman2001A, Kaufman2001B, Kaufman1995, KaufmanJTBspecial}, since feedback circuits provide a framework to understand basic dynamical features such as multi-stationarity, homeostasis and memory. The TR systems with their associated feedback circuits have been extended to the study of the fundamentals of the emergence of complex behaviour in simple-circuit structures in complex systems at large \cite{Kaufman2003complex, KaufmanJTBspecial}. The main results of this approach can be summarised in 3 general statements: (i) a positive circuit is necessary for the system to have stable states, (ii) a negative circuit is necessary for the system to exhibit, robust, sustained oscillations, and (iii) a necessary condition for chaos is the presence of both a positive and a negative circuit in the system. In particular, two classes of systems that R. Thomas and O. R\"{o}ssler called ``Labyrinth chaos'' and ``Arabesques'' \cite{ThomRoss2004, Arabesque2013, BasAnt2019} drew attention as examples of ``elegant chaos''. This signifies a type of minimal, generic, dynamical systems that highlight interesting peculiarities of chaotic behaviour \cite{SprottBook}.

The TR class of systems exhibits the whole repertoire of dynamics such as periodicity, multi-stationarity, coexisting attractors, bifurcation scenaria, etc. Furthermore, such systems with symmetric unstable stationary points, termed ``Labyrinth chaos" \cite{Thomas1999, ThomRoss2004, ThomRoss2006}, possess a peculiar, special, case of a chaotic state that occurs in deterministic systems in the absence of attractors, where the trajectories resemble fractional Brownian motion, termed ``Labyrinth walks'' (see Fig. \ref{fig:Labys}(c)). These trajectories were found to give rise to very interesting, symbolic dynamics \cite{SprottChlouverakis}. Interestingly, it has been reported recently that coupled arrays of ``Labyrinth chaos'' systems are able to exhibit stereotypical, chimera-like states \cite{BasAnt2019}, reminiscent of chimera states found in coupled Kuramoto oscillators \cite{Kuramotoetal2002,Abramsetal2004,Panaggioetal2015,BasAnt2019} and in other systems \cite{Hizanidisetal2016,Parasteshetal2018,Schmidtetal2017,Sigalasetal2018}.

\section{Labyrinth chaos and Labyrinth walks}

\begin{figure}[!h]
\centering
\begin{subfigure}{0.48\textwidth} 
\centering
\includegraphics[width=0.75\textwidth]{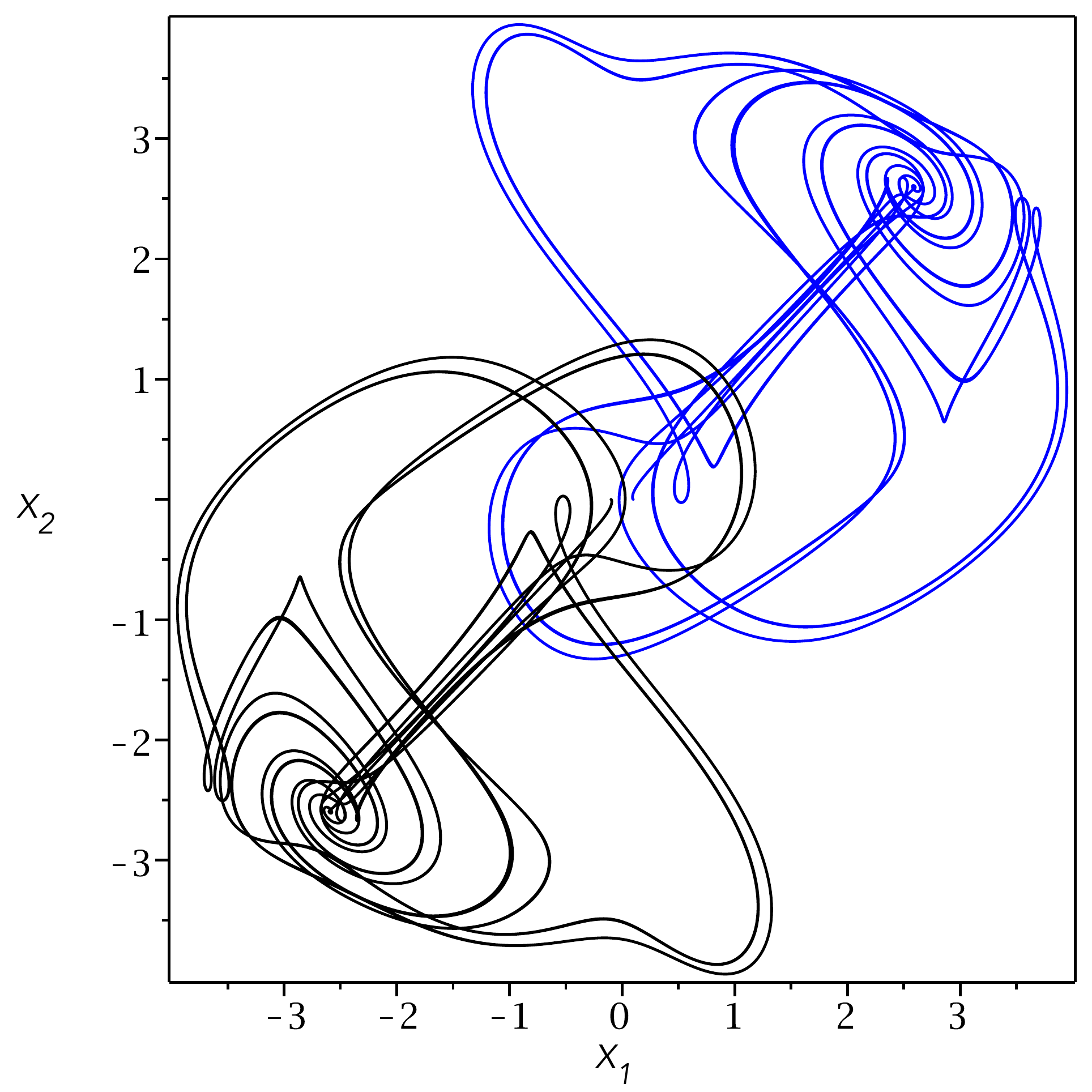}
\caption{}
\end{subfigure}
\begin{subfigure}{0.48\textwidth} 
\centering\includegraphics[width=0.75\textwidth]{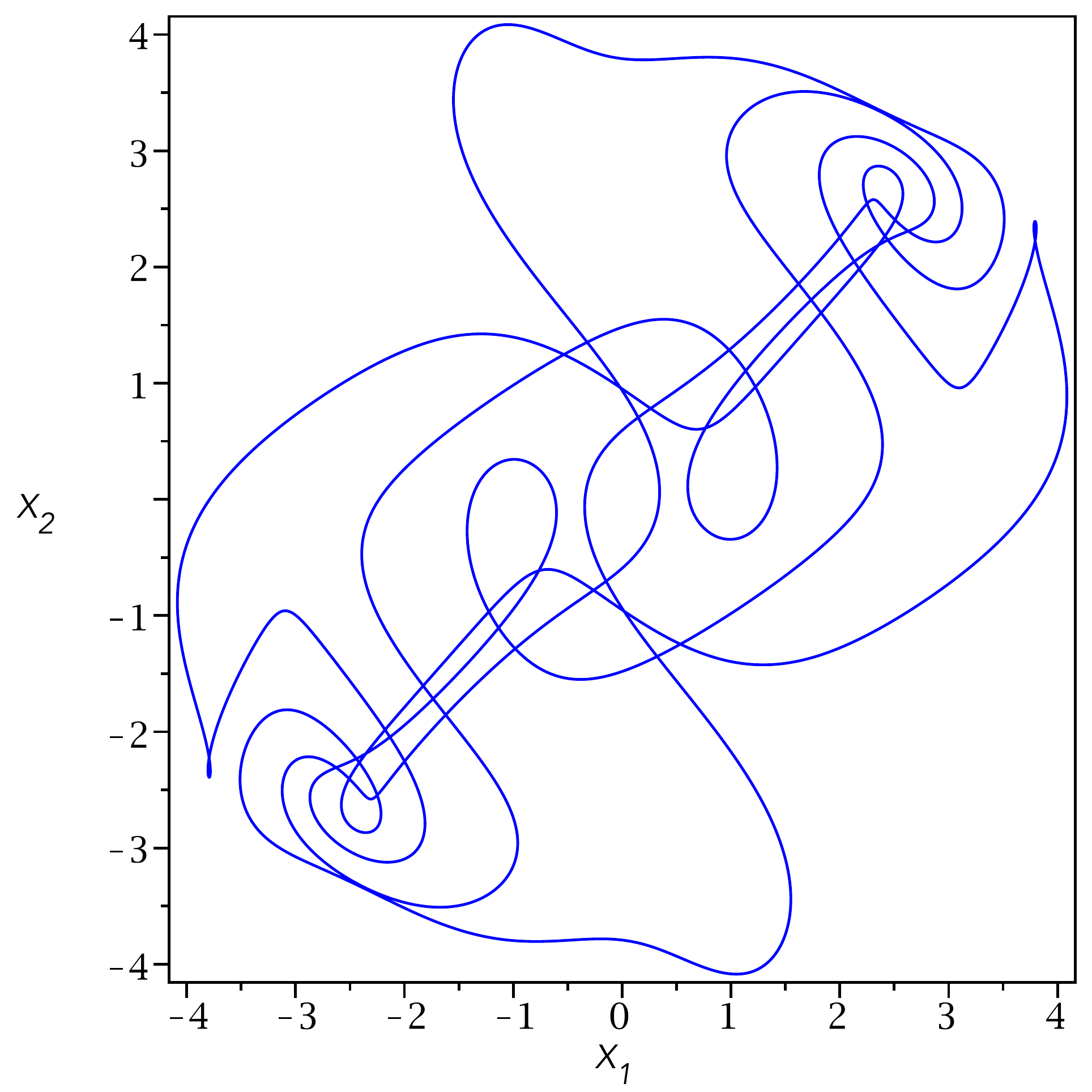}
\caption{}
\end{subfigure}
\begin{subfigure}{0.48\textwidth} 
\centering\includegraphics[width=0.75\textwidth]{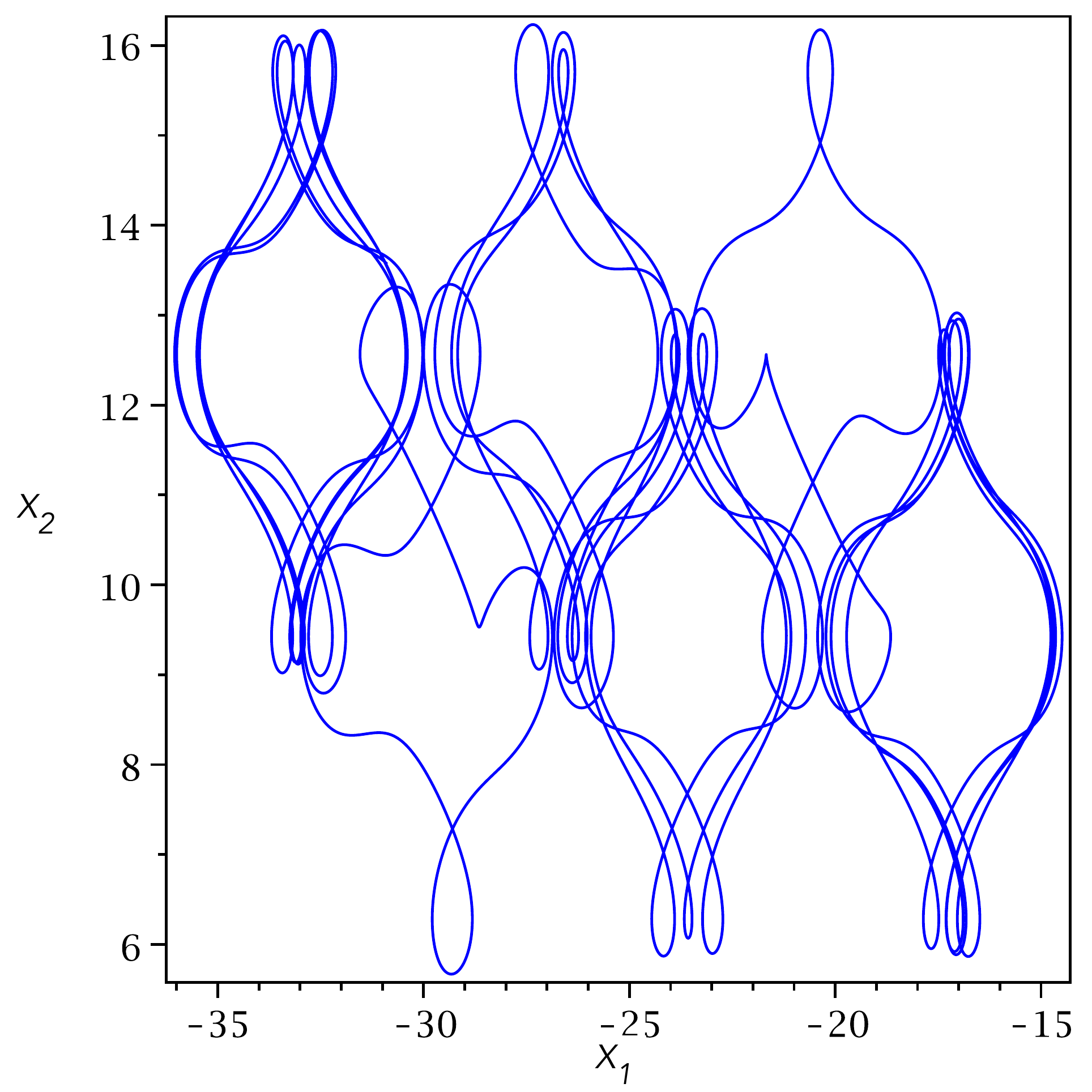}
\caption{}
\end{subfigure}
\caption{Typical trajectories of the $N=3$ TR system of Eq. \eqref{Aeq2}, projected on the ($X_1,X_2$)-plane with $X_3=0$, for 3 dissipation values $b$. Panel (a) is for $b=0.2$, showing coexisting chaotic attractors for 2 trajectories in blue and black. Panel (b) is for $b=0.19$ showing a complex periodic orbit. Panel (c) is for $b=0$ (i.e., conservative case), showing a ``Labyrinth walk''. Note that in this figure we have used a transient time of $t=400$ after which we plot the trajectories to disregard transient phenomena \cite{ChlouverakisSprott2007}.}\label{fig:Labys}
\end{figure}

The class of TR systems can be expressed as a set of $N$, cyclically coupled, ordinary differential equations
\begin{equation}\label{Aeq0}
\frac{dX_i}{dt} = -b X_i + F(X_{i+1})\quad\mod{N},
\end{equation}
where $X_i,\;i=1,2,\dots, N$ are real variables with $X_{N+1}=X_1$, a relation that introduces a periodic boundary condition to the system. The dissipation parameter $b$ serves as the only bifurcation parameter and, over the years, a variety of functions $F$ has been studied \cite{Thomas1999, ThomRoss2004, Arabesque2013, BasAnt2019}. This class of systems exhibits a repertoire of behaviour ranging from trivial stationary points to simple periodicity, to complicated periodicity, to chaos with coexisting strange attractors and bifurcation scenaria \cite{SprottChlouverakis,ChlouverakisSprott2007,BasAnt2019}. In Fig. (\ref{fig:Labys}), we present typical trajectories of the $N=3$ TR system of Eq. \eqref{Aeq0} with $F(x)=\sin(x)$ (see system \eqref{Aeq2}), projected on the ($X_1,X_2$)-plane with $X_3=0$, for 3 values of the dissipation parameter $b$. Panel (a) shows coexisting chaotic attractors for 2 trajectories depicted in blue and black, for $b=0.2$. For the relatively smaller value of $b=0.19$ in panel (b), one can see a complex periodic orbit. Finally, panel (c) shows for $b=0$ (i.e., the conservative case), a ``Labyrinth walk'', where only unstable stationary points exist in the phase space, as we show below.

Originally, drawing from its biological background, $F$ in Eq. \eqref{Aeq0} was considered as a model of threshold-type function. Interestingly though, it can also be a periodic function \cite{ThomRoss2004} and this is the case where novel dynamical behaviour can be observed when the dissipation term vanishes, i.e., for $b=0$ \cite{ThomRoss2004,ChlouverakisSprott2007,Arabesque2013, BasAnt2019}. Thus, in the case where $N=3$, $F(x)=\sin(x)$ and $b\in\mathbb{R}$, system \eqref{Aeq0} becomes
\begin{equation}\label{Aeq2}
\begin{aligned} 
\frac{d X_1}{d t} &=-b X_1+\sin (X_2),\\ 
\frac{d X_2}{d t} &=-b X_2+\sin (X_3),\\ 
\frac{d X_3}{d t} &=-b X_3+\sin (X_1).
\end{aligned}
\end{equation}
In the $N$-dimensional case, the trace of its Jacobian matrix (the rate of state-space contraction) is equal to $-bN$ and hence, when $b>0$ it has bounded solutions. Consequently, for $b=0$, it is conservative and the necessary condition (iii) for chaos is satisfied (see Introduction), however, no attractors are present! System \eqref{Aeq2} with $b=0$ is the so-called ``Labyrinth walks'' system (see also system \eqref{lw_3D_system}). In fact, in this case, the system possesses a countably infinite set of stationary points arranged in a $N$-dimensional lattice, where $\sin(X_i) =0,\;i=1,\dots, N$, i.e., the stationary points are given by $X_i=k\pi,\;k\in\mathbb{N}$. Amazingly, these are the only stationary points of the system and they all turn out to be unstable (what amounts to no attractors in the system), as when computing the eigenvalues $\lambda$ of the Jacobian matrix of system \eqref{Aeq2}, one obtains \cite{BasAnt2019}
\begin{equation*}
(-\lambda)^3=\prod_{i=1}^{3}\cos(X_i),\;X_i=k\pi,\;k\in\mathbb{N},
\end{equation*}
from which it results that
\begin{equation*}
\lambda^3=\left\{
\begin{array}{ll}
1\mbox{ or}\\
-1\\
\end{array},
\right.
\end{equation*}
depending on the values of $X_i$. For $\lambda^3=1$, one obtains
\begin{equation*}
\left\{
\begin{array}{ll}
\lambda_1=1,\mbox{ or }\\
\lambda_2=-\frac{1}{2} -\frac{\sqrt{3}}{2} i,\mbox{ or }\\
\lambda_3=-\frac{1}{2} +\frac{\sqrt{3}}{2} i,\\
\end{array} 
\right.
\end{equation*}
from which it is concluded that the stationary points $X_i=k\pi,\;k\in\mathbb{N}$ are unstable since $\lambda_1>0$ and similarly, for $\lambda^3=-1$, one obtains
\begin{equation*}
\left\{
\begin{array}{ll}
\lambda_1=-1,\mbox{ or }\\
\lambda_2=\frac{1}{2} -\frac{\sqrt{3}}{2} i,\mbox{ or }\\
\lambda_3=\frac{1}{2} +\frac{\sqrt{3}}{2} i,\\
\end{array} 
\right.
\end{equation*}
from which, again, it is concluded that the stationary points $X_i=k\pi,\;k\in\mathbb{N}$ are unstable since this time the real part of $\lambda_{2,3}$, $\Re(\lambda_{2,3})=\frac{1}{2}>0$.

As mentioned earlier, system \eqref{Aeq2} is conservative for $b=0$ (``Labyrinth walks'' system) and, thus, conserves phase space volumes in the sense of Liouville's theorem. It is thus tempting to explore whether it is also a Hamiltonian system. As we show in the next section, this is not the case at all! It is worth it to mention that for $b=0$, it is time-reversible ($t\rightarrow -t$) and it preserves parity, i.e., $\{X_i\}\rightarrow\{-X_i\}$ for $i=1,\ldots,N$, while depending on $N$, a plethora of symmetries is present. The simplest possible case of ``Labyrinth chaos'' is for $N=3$, i.e., the system comprises the minimal number of ordinary differential equations necessary to exhibit chaos \cite{ChlouverakisSprott2007,BasAnt2019}.

A physical phenomenon that can be described by such a system could be a particle moving in an $N$-dimensional lattice or ``hyperlabyrinth'' under the influence of some external source of energy \cite{ChlouverakisSprott2007}. Surprisingly, the equations in system \eqref{Aeq2} are similar to the Ikeda delay differential equation \cite{Ikedaetal1987}
\begin{equation*}
\frac{dx(t)}{dt}=-bx(t)+\sin(x(t-\tau)),
\end{equation*}
where $\tau$ is the time delay. This system is known to result in high-dimensional chaos and is used for modeling lasers with wavelength hyperchaos \cite{Udaltsovetal2001,ChlouverakisSprott2007}. Its behaviour is, thus, reminiscent of the behaviour exhibited by the ``Labyrinth chaos'' system \eqref{Aeq2}.

In the next section, we will focus on system \eqref{Aeq2} with $b=0$ (i.e., on the ``Labyrinth walks'' system) and will show that even though it is conservative, it does not admit an autonomous Hamiltonian as a constant function in time. As a consequence, the system does not admit a symplectic structure either. This interesting combination of properties makes it an elegant example of a chaotic, conservative, non-Hamiltonian system, with no attractors in its phase space, leading to trajectories resembling fractional Brownian motion \cite{Thomas1999, ThomRoss2004, ThomRoss2006}, occurring though in a deterministic system!

\section{``Labyrinth walks'' does not admit a Hamiltonian}

Here, we show that system \eqref{Aeq2} with $b=0$, i.e., the ``Labyrinth walks'' system
\begin{equation}\label{lw_3D_system}
\begin{aligned} 
\frac{d X_1}{d t} &=\sin (X_2),\\ 
\frac{d X_2}{d t} &=\sin (X_3),\\ 
\frac{d X_3}{d t} &=\sin (X_1),
\end{aligned}
\end{equation}
cannot admit an autonomous Hamiltonian $H$ \cite{Hamiltonian_Torrealdea} and thus, a symplectic structure. We note that system \eqref{lw_3D_system} is autonomous.

In particular, an autonomous system $\dot{X}=f(X)$ is Hamiltonian if it has the form
\begin{equation}\label{skew_symmetric_J}
\dot{X}=J\nabla{H(X)},
\end{equation}
where $X\in\mathbb{R}^{2N}$, $H(X)$ denotes the Hamiltonian function, and
\begin{equation*}
J=
\left( {\begin{array}{cc}
0 & I_d \\
-I_d & 0 \\
\end{array} } \right),
\end{equation*}
where $I_d$ denotes the $N\times N$ identity matrix in $\mathbb{R}^N$. As $J$ is a skew-symmetric matrix that satisfies the Jacobi's closure condition \cite{Olver1991}, $\dot{X}$ and $H(X)$ are always orthogonal. Consequently, motion in the phase space takes place at a constant value of the Hamiltonian function, that is, $H(X)$ is a first integral of motion and the system $\dot{X}=f(X)$ is conservative. Here, we show that even though, system \eqref{lw_3D_system} is conservative, it is not Hamiltonian.

To show this, we start with a general, $N=3$-dimensional, autonomous, conservative dynamical system $\dot{X}=f(X)$ given by \cite{Sarasolaetal2004}
\begin{equation}\label{Aeq1}
\nabla H^Tf(X)=0,
\end{equation}
where $H(X)$ is the Hamiltonian with $X=(X_1,X_2,X_3)$, $f$ a smooth function from $\mathbb{R}^3$ to $\mathbb{R}^3$ and $\nabla H^T$ is the transpose of $\nabla H=\bigl(\frac{\partial H}{\partial X_1},\frac{\partial H}{\partial X_2},\frac{\partial H}{\partial X_3}\bigr)$. To find the Hamiltonian $H(X)$ of the conservative system $\dot{X}=f(X)$, one needs to find the function $H(X)$ that satisfies Eq. \eqref{Aeq1} (see also \cite{Kobe}). If that is possible, then $H(X)$ will be a constant function in time, that is an integral of motion of the system $\dot{X}=f(X)$.

Furthermore, if $H$ exists, it should satisfy Eq. \eqref{skew_symmetric_J}. In the case of system \eqref{lw_3D_system}
\begin{equation}\label{f_c}
f(X)=\left(\begin{array}{c}{\sin (X_2)}\\ {\sin (X_3)}\\ {\sin (X_1)}\end{array}\right).
\end{equation}

Substituting Eq. \eqref{f_c} to Eq. (\ref{Aeq1}), we obtain
\begin{equation}\label{Aeq4}
\frac{\partial H}{\partial X_1}\sin (X_2) +
\frac{\partial H}{\partial X_2}\sin (X_3) +
\frac{\partial H}{\partial X_3}\sin (X_1) = 0.
\end{equation}
One has to find $H(X)$ that satisfies this partial differential equation, if system \eqref{lw_3D_system} can admit a Hamiltonian $H$. We will show below that this is not the case for system \eqref{lw_3D_system}.

Let's suppose that ${\partial H}/{\partial X_i}\neq 0,\;i=1,2,3$ as, obviously, if they are all equal to zero, then $H(X)$ will be identically equal to a constant and we are looking for a non-identically equal to a constant function $H$. Focusing on Eq. \eqref{Aeq4}, we notice there are infinitely many choices for ${\partial H}/{\partial X_1}$, ${\partial H}/{\partial X_2}$ and ${\partial H}/{\partial X_3}$ consistent with Eq. \eqref{lw_3D_system}, since they are assumed different than zero. Here, we elucidate on two such choices as they can help us demonstrate why none of these infinitely many choices are actually compatible with Eq. \eqref{Aeq4}, leading to the conclusion that it is not possible to find a Hamiltonian $H$ that can satisfy it.

The first option, which is one of the simplest, is given by the set of equations
\begin{subequations}\label{Aeq5}
\begin{align}
\frac{\partial H}{\partial X_1}&=g(X_1, X_2, X_3)\sin (X_3),\\ 
\frac{\partial H}{\partial X_2}&=-g(X_1, X_2, X_3)\sin (X_2)-h(X_1, X_2, X_3)\sin (X_1),\\ 
\frac{\partial H}{\partial X_3}&=h(X_1, X_2, X_3)\sin (X_3),
\end{align}
\end{subequations}
where $g$ and $h$ are smooth functions in ${\mathbb R}^3$ such that $g$ does not  contain explicitly the terms $\sin(X_2)$ and $\sin(X_3)$, and $h$ the terms $\sin(X_1)$ and $\sin(X_3)$. These conditions reflect the fact that the variables in Eq. (\ref{Aeq0}) are cyclic.
 
The second option we study here is given by the set of equations
\begin{equation}\label{Aeq6}
\begin{aligned}
\frac{\partial H}{\partial X_1}&=g_{1}(X_1, X_2, X_3)\sin (X_3)-g_{2}(X_1, X_2, X_3)\sin (X_1),\\
\frac{\partial H}{\partial X_2}&=-g_{1}(X_1, X_2, X_3)\sin (X_2)-h_{1}(X_1, X_2, X_3)\sin (X_1),\\
\frac{\partial H}{\partial X_3}&=h_{1}(X_1, X_2, X_3)\sin (X_3)+g_{1}(X_1, X_2, X_3)\sin (X_2),
\end{aligned}
\end{equation}
where $g_1$, $g_2$ and $h_1$ are smooth functions in ${\mathbb R}^3$ such that $g_1$ does not contain explicitly the terms $\sin(X_2)$ and $\sin(X_3)$, $g_2$ the terms $\sin(X_1)$ and $\sin(X_2)$, and $h_1$ the terms $\sin(X_1)$ and $\sin(X_3)$.

Let us first consider the case of Eqs. \eqref{Aeq5}. Integrating Eq. (\ref{Aeq5}a) with respect to $X_1$ yields
\begin{equation}\label{Aeq7}
H(X_1, X_2, X_3)=\sin (X_3)\int g(X_1, X_2, X_3) dX_1+A(X_2, X_3),
\end{equation}
where $A(X_2,X_3)$ plays the role of the ``constant'' of integration. Computing ${\partial H}/{\partial X_2}$ using Eq. \eqref{Aeq7} and comparing it to Eq. (\ref{Aeq5}b), yields
\begin{equation}\label{Aeq8}
\frac{\partial A(X_2, X_3)}{\partial X_2}=-\sin (X_3)\int\frac{\partial g(X_1,X_2,X_3)}{\partial X_2} d X_1-g(X_1, X_2, X_3)\sin (X_2)-h(X_1, X_2, X_3)\sin (X_1).
\end{equation}

On the left hand side of Eq. \eqref{Aeq8}, there is no dependence on $X_1$, while on the right hand side, even if we suppose that $g$ and $h$ depend on $X_2$ and $X_3$ only, we cannot eliminate its dependence on $X_1$ because of the term $\sin(X_1)$. Thus, there are two possibilities for Eq. \eqref{Aeq4} to hold. The first would be to set $h = 0$, but then according to Eq. (\ref{Aeq5}c), ${\partial H}/{\partial X_3} = 0$, which contradicts the initial hypothesis! The second possibility would be to set ${\partial A}/{\partial X_3}=0$. Then, the derivative of Eq. \eqref{Aeq8} with respect to $X_1$ yields
\begin{equation}\label{Aeq9}
\frac{\partial g}{\partial X_2}\sin(X_3)+\frac{\partial g}{\partial X_1}\sin(X_2)+\frac{\partial}{\partial X_1}\bigr(h\sin (X_1)\bigl)=0.
\end{equation}
Following the same procedure, Eqs. (\ref{Aeq5}b) and (\ref{Aeq5}c) yield
\begin{equation}\label{Aeq10}
\frac{\partial h}{\partial X_2}\sin (X_3)+\frac{\partial g}{\partial X_3}\sin (X_2)+\frac{\partial h}{\partial X_3}\sin (X_1)=0.
\end{equation}
However, we have assumed that $g$ does not contain the terms $\sin(X_2)$ and $\sin(X_3)$, and that $h$ does not contain the terms $\sin(X_1)$ and $\sin(X_3)$. Thus, the nullity of Eqs. \eqref{Aeq9} and \eqref{Aeq10} becomes impossible.

It is now easy to see why the second option in Eqs. \eqref{Aeq6} or any other option for ${\partial H}/{\partial X_1}$, ${\partial H}/{\partial X_2}$ and ${\partial H}/{\partial X_3}$ consistent with Eq. \eqref{Aeq4} will result to similar contradictions.

Thus, the only remaining case is when one of the partial derivatives ${\partial H}/{\partial X_1}$, ${\partial H}/{\partial X_2}$ and ${\partial H}/{\partial X_3}$ in Eq. \eqref{Aeq4} is equal to 0. For example, let us assume that ${\partial H}/{\partial X_1}=0$. Then, Eq. \eqref{Aeq4} becomes 
\begin{equation}\label{Aeq11}
\frac{\partial H}{\partial X_2}\sin(X_3) +\frac{\partial H}{\partial X_3}\sin(X_1)=0.
\end{equation}
However, having assumed that ${\partial H}/{\partial X_1}=0$, it means that $H$ does not depend on $X_1$, which contradicts Eq. \eqref{Aeq11}, since again the dependence on $X_1$ due to the term $\sin(X_1)$ cannot be eliminated. Hence, we conclude that it is not possible to find an autonomous Hamiltonian function $H(X_1, X_2, X_3)$ that would be a solution to the partial deferential equation \eqref{Aeq1} (see also Eq. \eqref{Aeq4}) and, thus, system \eqref{lw_3D_system} is not Hamiltonian. Consequently, it does not admit a symplectic structure either, as there is no $H$  that can satisfy Eq. \eqref{skew_symmetric_J}.

It is worth it to note, that even though, system \eqref{lw_3D_system} is not Hamiltonian, it has a vector potential as $\nabla f=0$ for $f$ in Eq. \eqref{f_c}. Thus, there exists a vector field $F(F_1,F_2,F_3 )$, called the vector potential \cite{vectorpot}, such that $\nabla\times F=f$, where $\times$ is the cross-product. This results to the following system of partial differential equations
\begin{equation*}
\begin{aligned} 
\frac{\partial F_{3}}{\partial X_2}-\frac{\partial F_{2}}{\partial X_3}&=\sin (X_2),\\
\frac{\partial F_{1}}{\partial X_3}-\frac{\partial F_{3}}{\partial X_1}&=\sin (X_3),\\
\frac{\partial F_{2}}{\partial X_1}-\frac{\partial F_{1}}{\partial X_2}&=\sin (X_1),
\end{aligned}
\end{equation*}
from which it follows that $F$ is not unique. For example, to find a simple solution, we can set $F_3 = 0$. Straight forward calculations then yield to the conclusion that $F$ should be of the form
\begin{equation*}
F(-\cos(X_3),-X_3\sin(X_2)-\cos(X_1),0),
\end{equation*}
but that is only one possibility. For a conservative system in general, a vector potential $F$ is related to the flow of the vector field $f$ through Stokes' theorem.

\section{Conclusions \& Outlook}

The conservative version of ``Labyrinth chaos'', the so-called ``Labyrinth walks'' (see Eq. \eqref{lw_3D_system}), in all its simplicity and elegance, suggests the possibility of being given by a Hamiltonian function, as is known for other nonlinear, cyclically coupled, systems, such as the Lotka-Voltera system and its variants \cite{LVgen}, the Arnold-Beltrami-Childress (``1:1:1 ABC'') model \cite{ABC111a,ABC111b}, etc. As we have shown here, it is not possible to find such a Hamiltonian function \cite{Kobe, Olver1991, Hamiltonian_Torrealdea} for system \eqref{lw_3D_system}.  As a consequence, it does not admit a symplectic structure. However, it is conservative, and thus admits a vector potential, being at the same time chaotic. This combination of properties, makes it an elegant example of a chaotic, conservative, non-Hamiltonian system, with no attractors in its phase space, leading to trajectories resembling fractional Brownian motion \cite{Thomas1999, ThomRoss2004, ThomRoss2006}, occurring though in a deterministic system!

There is a class of conservative systems in the context of Eq. \eqref{Aeq1}, which are not Hamiltonian due to specific considerations pertaining to the construction of ``Hoover thermostats'' in Statistical Mechanics \cite{HooverA,HooverB}. Although this theory provides a framework for the analysis of non-Hamiltonian systems and for determining the phase space distribution generated by the equations of motion assuming ergodicity, the kinetic, dissipative and potential parts are specific in expressing a balance that makes the phase space incompressible. Still, the connection to chaotic dynamics has not been reported or elucidated.

Finally, the properties of ``Labyrinth chaos'', including those of ``Labyrinth walks'', places it at a class of its own. Following the work in this paper, there are three main points to consider in future work: (i) the discrepancy, as far as a Hamiltonian structure is concerned, between the right hand side of system \eqref{Aeq2} and its approximation via Taylor expansions, what gives rise to ``Arabesque''-like systems, especially to systems with $b=0$ \cite{ThomRoss2006,Arabesque2013}, (ii) the differences in symmetry considerations and in kinetic versus potential terms in Eqs. \eqref{Aeq0} and \eqref{Aeq1}, and the ``(1:1:1)ABC'' model \cite{ABC111a, ABC111b}, where combinations of only ``$\sin(\cdot)$'' versus ``$\sin(\cdot)$'' and ``$\cos(\cdot)$'' terms appear in their right hand sides, and (iii) the role of the vector potential as a ``carrier of information'' \cite{Hiley2002,Hiley1999,vectorpot}, which is an exciting field of research to pursue further, especially, in connection to arrays of ``Labyrinth chaos'' systems, where chimera-like states have been shown to emerge \cite{BasAnt2019}.

\section{Acknowledgements}
This research was supported by our respective Institutes: Department of Physics, Qom University of Technology for AL, Service de Physique des Syst\`emes Complexes et M\'ecanique Statistique and CeNoLi, Universit\'e Libre de Bruxelles for VB and Department of Mathematical Sciences, University of Essex for CGA.


\end{document}